\documentclass[twocolumn,showpacs,preprintnumbers,amsmath,amssymb]{revtex4}
\usepackage{graphicx}
\usepackage{dcolumn}
\usepackage{bm}

\begin{document}

\draft
\title{Dilatonic wormholes:
construction, operation, maintenance and collapse to black holes}
\author{Sean A. Hayward}\email{hayward@mm.ewha.ac.kr}
\author{Sung-Won Kim}\email{sungwon@mm.ewha.ac.kr}
\author{Hyunjoo Lee}\email{hyunjoo@mm.ewha.ac.kr}
 \affiliation{Department of Science Education, Ewha Womans University, Seoul
120-750, Korea}

\date{\today}

\begin{abstract}
The CGHS two-dimensional dilaton gravity model is generalized to include a
ghost Klein-Gordon field, i.e.\ with negative gravitational coupling. This
exotic radiation supports the existence of static traversible wormhole
solutions, analogous to Morris-Thorne wormholes. Since the field equations are
explicitly integrable, concrete examples can be given of various dynamic
wormhole processes, as follows. (i) Static wormholes are constructed by
irradiating an initially static black hole with the ghost field. (ii) The
operation of a wormhole to transport matter or radiation between the two
universes is described, including the back-reaction on the wormhole, which is
found to exhibit a type of neutral stability. (iii) It is shown how to maintain
an operating wormhole in a static state, or return it to its original state, by
turning up the ghost field. (iv) If the ghost field is turned off, either
instantaneously or gradually, the wormhole collapses into a black hole.
\end{abstract}

\maketitle

\section{Introduction}

The theoretical existence of space-time wormholes has intrigued experts and
public alike. Wheeler \cite{W} speculated that quantum fluctuations in
space-time topology can occur, so that the smooth space-time of classical
Einstein gravity becomes, at the Planck scale, a continual foam of short-lived
interconnections. If such a wormhole could be expanded to macroscopic size, it
could provide a short cut between otherwise distant regions, like a bookworm
tunnelling between different pages of an atlas. Certainly all the standard
stationary black-hole solutions have wormhole ($R\times S^2$) spatial topology,
describing two Alice-through-the-looking-glass universes joined by the famous
Einstein-Rosen bridge \cite{ER}, or wormhole throat. In such cases, the two
universes are not causally connected, so that it is impossible to travel
between them, any attempt leading instead into the black hole \cite{K}.
However, this is only just so; a light ray can be sent along the past boundary
of one universe to the future boundary of the other, just escaping the black
hole. Thus it is not difficult to imagine a small modification which would
yield a traversible wormhole.

Morris \& Thorne \cite{MT} popularized traversible wormholes as a respectable
theoretical possibility, studying static, spherically symmetric cases in
detail. The spatial topology is the same as in the black-hole cases, but the
throat or minimal surface is preserved in time, so that observers can pass
through it in either direction, travelling freely between the two universes.
According to the Einstein equation, such a geometry requires matter which does
not satisfy a classical positive-energy property, the weak energy condition.
However, this condition can be violated quantum-mechanically, e.g.\ in the
Casimir effect, or in alternative gravity theories. Such negative-energy matter
was dubbed exotic matter by Morris \& Thorne. This provoked renewed interest in
traversible wormholes, as reviewed by Visser \cite{V}. The possible
astrophysical existence of wormholes has been taken seriously enough for
searches of observational data \cite{TRA}.

The Morris-Thorne framework begs development in at least two important
respects, quite apart from generalizing beyond static, spherically symmetric
cases. Firstly, the exotic matter was not modelled, but simply assumed to exist
in exactly the configuration needed to support the wormhole. Secondly, the
back-reaction of the transported matter on the wormhole was ignored. If the
wormhole turns out to be unstable to such back-reaction, it would be useless
for actual transport. Many physicists' instinctive reaction is that negative
energy, unbounded below, will lead to instability. Another practical question
is: if the negative-energy source fails, i.e.\ the exotic matter is not
maintained, what happens to the wormhole? Again, a pessimistic reaction is that
the negative energy densities are likely to create naked singularities. An
alternative prediction was that the wormhole would collapse to a black hole
\cite{wh}. The same reference introduced a framework for studying dynamic
wormholes, in which both wormholes and black holes are locally defined by the
same geometrical objects, trapping horizons, with one key difference: black
holes have achronal horizons and wormholes have temporal horizons, so that they
are respectively one-way and two-way traversible. This also indicated that a
wormhole could be constructed from a black hole using exotic matter.

To find definite answers to the above questions, it is necessary to specify the
exotic matter model. Various studies have been based on alternative gravity
theories or semi-classical quantum field theory, e.g.\ \cite{HPS,KL}. However,
the difficulty of solving the field equations tends to obscure issues of
principle. Here we propose a toy model, intended to describe the essential
dynamics of a wormhole, but explicitly integrable, so that concrete answers to
the above questions are easily found. A similar approach in the context of
black-hole evaporation involved the CGHS two-dimensional dilaton gravity model
\cite{CGHS}, which was expected to share similar features with spherically
symmetric black-hole evaporation by Hawking radiation. Certainly classical
black-hole dynamics is qualitatively similar, for instance in regard to cosmic
censorship, but such properties are much easier to prove in the toy model
\cite{cc} than in the corresponding realistic model, the spherically symmetric
Einstein-Klein-Gordon system \cite{C}. Here we propose a simple generalization
of the CGHS model to include a ghost Klein-Gordon field, i.e.\ with the
gravitational coupling taking the opposite sign to normal. This is a specific
model for the exotic matter, or more accurately exotic radiation, as we take
the massless field for simplicity.

The article is organised as follows. \S II describes the model and how its
general solution can be constructed from initial data. \S III reviews the
static black-hole solution and introduces the static wormhole solution, each of
which depend on one parameter, mass or horizon radius. \S IV describes what
happens to an initially static wormhole when the ghost field supporting it is
turned off and allowed to disperse. \S V shows how to construct a static
wormhole by irradiating an initially static black hole with the ghost field. \S
VI describes the operation of a wormhole for transport or signalling, using a
normal Klein-Gordon field to model the matter or radiation, including the
back-reaction of the field on the wormhole. \S VII shows how to maintain the
operating wormhole in a static state, or return it to its original state. \S
VIII concludes.

\section{Dilaton gravity}

The CGHS two-dimensional dilaton gravity of Callan {\it et al.} \cite{CGHS} is
generalized by the action
\begin{equation}\label{action}
 \int_S \mu
\left[ e^{-2\phi } \left( R + 4 (\nabla \phi )^2  + 4\lambda ^2 \right) -
\frac{1}{2} (\nabla f)^2
 + \frac{1}{2}(\nabla g)^2  \right]
\end{equation}
where $S$ is a 2-manifold, $\mu$, $R$ and $\nabla$ are the area form, Ricci
scalar and covariant derivative of a Lorentz 2-metric on $S$, $\lambda$
represents a negative cosmological constant, $\phi$ is a scalar dilaton field,
$f$ is a scalar field representing matter and $g$ is a ghost scalar field with
negative gravitational coupling. The last term is added to the CGHS action in
order that g provides the negative energy densities needed to maintain a
wormhole \cite{MT,V,wh,HV1,HV2,IH,wd}.

\begin{figure}
\includegraphics[width=9cm,height=13cm,angle=0]{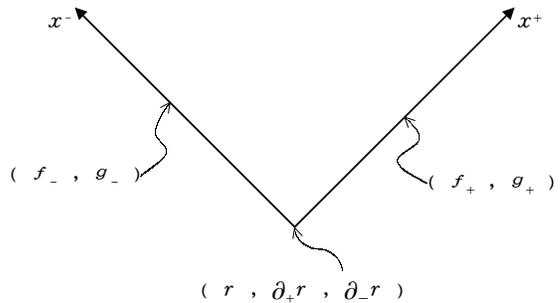}
 \vspace{-80mm}
 \caption{Initial data, taking $x^\pm_0=0$.} \label{initialdata}
\end{figure}

By choosing future-pointing null coordinates $(x^+ , x^- )$, the line element
may be written as
\begin{equation}\label{metric}
  ds^2 = -2 e^{2\rho} dx^+ dx^-
\end{equation}
where the conventional factor of 2 differs from that of earlier references
\cite{CGHS,cc}. One component of the field equations is
\begin{equation}\label{dilaton eqn}
  \partial_+ \partial_- \phi = \partial_+ \partial_- \rho
\end{equation}
where $\partial_\pm=\partial/\partial x^\pm$, so the coordinate
freedom $x^\pm\mapsto\hat x^\pm(x^\pm)$ can be used to take
\begin{equation}\label{fix}
\rho = \phi.
\end{equation}
The remaining coordinate freedom is just
\begin{equation}\label{freedom}
x^\pm\mapsto e^{\pm b}x^\pm+c^\pm
\end{equation}
where the constants $c^\pm$ fix the origin and the constant $b$ refers to
relative rescalings of $x^\pm$. It is convenient to transform the dilaton field
$\phi$ to
\begin{equation}\label{r}
r=2e^{-2\phi}.
\end{equation}
Then the remaining field equations are the evolution equations
\begin{eqnarray}
\label{fequation}
 \partial_+ \partial_- f &=& 0 \\
 \label{gequation}
 \partial_+ \partial_- g &=& 0 \\
 \label{requation}
 \partial_+ \partial_- r &=& -4 \lambda^2
\end{eqnarray}
and the constraints
\begin{equation}\label{constraints0}
  \partial_\pm   \partial_\pm  r = ( \partial_\pm  g )^2  -
  (\partial_\pm f )^2 .
\end{equation}
The evolution equations (\ref{fequation}-\ref{requation}) have the
general solutions
\begin{eqnarray}
\label{fsolution}
  f(x^+,x^-) &=& f_+ (x^+) + f_- (x^-) \\
  \label{gsolution}
  g(x^+,x^-) &=& g_+ (x^+) + g_- (x^-) \\
  \label{rsolution}
  r(x^+,x^-) &=& r_+ (x^+) + r_- (x^-) - 4\lambda ^2 x^+ x^-.
\end{eqnarray}
The constraints (\ref{constraints0}) are preserved by the evolution equations
in the $\partial_\mp$ directions, and so may be reduced to
\begin{equation}\label{constraints}
  \partial_\pm \partial_\pm r_\pm = {G_\pm}^2 - {F_\pm}^2
\end{equation}
where the null derivatives
\begin{eqnarray}
F_\pm&=&\partial_\pm f_\pm\\
G_\pm&=&\partial_\pm g_\pm
\end{eqnarray}
are convenient variables. The constraints are manifestly integrable for $r_\pm$
given initial data
\begin{eqnarray}\label{data}
(f_+,g_+)\qquad&&\hbox{on $x^-=x_0^-$}\\
(f_-,g_-)\qquad&&\hbox{on $x^+=x_0^+$}\\
(r,\partial_+r,\partial_-r)\qquad&&\hbox{at $x^+=x_0^+$, $x^-=x_0^-$}
\end{eqnarray}
for constants $x_0^\pm$ (Fig.~\ref{initialdata}). Then the general procedure is
to specify this initial data, integrate the constraints (\ref{constraints}) for
$r_\pm$, then the general solution follows as
(\ref{fsolution}-\ref{rsolution}).

Finally, we note from the vacuum constraints (\ref{constraints0}) that $r$ is
analogous to the areal radius in spherically symmetric Einstein gravity
\cite{cc,1st}, correctly reproducing the expansion of a light wave in flat
space-time, $\partial_\pm\partial_\pm r=0$.

\section{Static black-hole and wormhole solutions}

\begin{figure}
\includegraphics[width=8cm,height=12cm,angle=0]{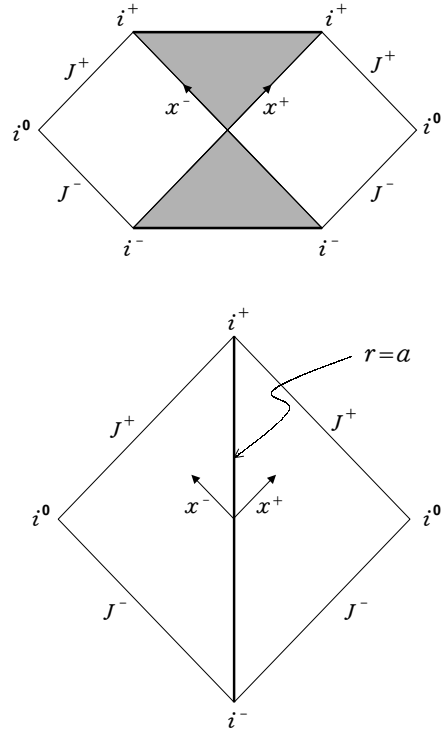}
\vspace{-15mm}
 \caption{Penrose conformal diagrams of (a) the static black hole,
 (b) the static wormhole.
 Both space-times are divided into two universes (unshaded regions),
 but observers can travel freely between them via the wormhole,
 whereas the black hole (upper shaded region) swallows any such attempt.}
 \label{staticbhwh}
\end{figure}

In vacuum, $f=g=0$, the general solution to the field equations is \cite{CGHS}
\begin{equation}\label{vacuumsol}
  r = 2m - 4\lambda^2  x^+ x^-
\end{equation}
where the origin has been fixed using (\ref{freedom}). The constant $m$ may be
interpreted as the mass of the space-time, whose global structure has been
described previously \cite{cc}. For $m>0$ there is a static black hole,
analogous to the Schwarzschild black hole (Fig.~\ref{staticbhwh}(a)). The case
$m=0$ is flat and the case $m<0$ contains an eternal naked singularity,
analogous to the negative-mass Schwarzschild solution. Henceforth we take
$m>0$. Note that the original reference \cite{CGHS} defined mass as $\lambda
m$, in which case $\lambda r$ would be analogous to radius, but for $\lambda>0$
this makes no essential difference in the following.

Next, we consider the case that $f=0$ but $g \neq 0 $, finding the solution
\begin{eqnarray}\label{staticwhsol}
  r &=& a + 2\lambda^2 (x^+ - x^- )^2 \\
  g &=& 2\lambda (x^+ - x^- )
\end{eqnarray}
where the origin has again been fixed. A coordinate transformation $x^\pm=(t\pm
z)/\sqrt{2}$ shows that the solution is static:
\begin{equation}
  ds^2 = \frac{2(dz^2-dt^2)}{a+4\lambda^2 z^2} .
\end{equation}
If $a>0$, this represents a traversible wormhole with analogous global
structure to a Morris-Thorne wormhole, with a throat $r=a$ at $z=0$, joining
two regions with $r>a$, a $z>0$ universe and a reflected $z<0$ universe
(Fig.~\ref{staticbhwh}(b)). If $a < 0$, there is an eternal naked singularity
$r=0$, while if $a=0$, the space-time has constant negative curvature, as can
be seen by calculating the Ricci scalar \cite{cc}
\begin{equation}
  R=r^{-1}\partial_+r\partial_-r -\partial_+\partial_-r.
\end{equation}
Henceforth we take $a>0$.

In summary, the model naturally contains both static black holes and static
traversible wormholes. Before proceeding, we note an important feature of both
cases, the trapping horizons, defined by $\nabla r\cdot\nabla r=0$, or
equivalently $\partial_+r=0$ or $\partial_-r=0$ \cite{wh,cc,1st,bhd}. In the
black hole, they coincide with the event horizons $r=2m$ at $x^-=0$ and $x^+=0$
respectively. In the wormhole, there is a double trapping horizon,
$\partial_+r=\partial_-r=0$, at the throat $r=a$. This illustrates how trapping
horizons of different type may be used to locally define both black holes and
wormholes \cite{wh}. Locating the trapping horizons, or equivalently the
trapped regions where $\nabla r\cdot\nabla r<0$, is a key feature of the
following analysis of dynamic situations.

\section{Wormhole collapse}

One can now study what happens to a static wormhole if its negative-energy
source fails. We consider first that the supporting ghost field $g$ is switched
off suddenly from both sides of the wormhole, then that $g$ is instead
gradually reduced to zero.

\subsection{Sudden collapse}

We set the initial data so that there is a static wormhole, with $g$ then
switched off suddenly from both sides of the wormhole (Fig.~\ref{sud}(a)):

\begin{figure}
\includegraphics[width=7cm,height=10cm,angle=0]{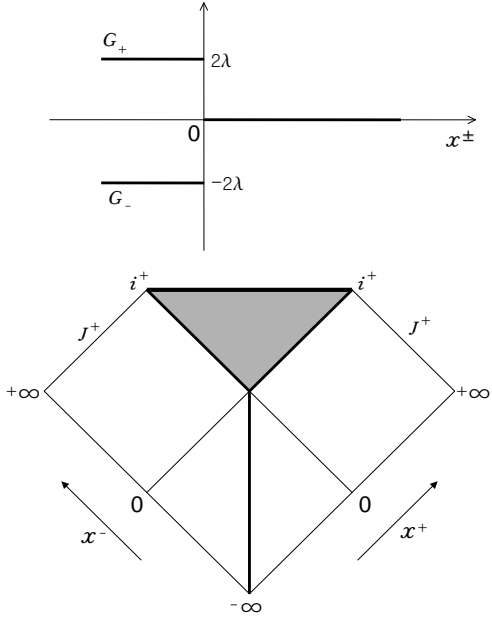}
 \vspace{-10mm}
 \caption{(a) The ghost field $g$ is switched off suddenly
from both sides of the wormhole. (b) Conformal diagram: when the wormhole's
negative-energy source fails suddenly at $x^\pm = 0$, it immediately collapses
into a black hole.}
 \label{sud}
\end{figure}

\begin{eqnarray}\label{suddenGeq}
G_\pm = \left\{
\begin{array}{ll}
\pm 2\lambda\,,  & x^\pm < 0 \\
 0\,, &x^\pm \geq 0.
\end{array}\right.
\end{eqnarray}
Taking $f=0$, the constraints (\ref{constraints}) are
\begin{eqnarray}\label{suddenreq}
  \partial_\pm   \partial_\pm  r_\pm =
  \left\{
\begin{array}{ll}
4 \lambda^2\,, &x^\pm < 0 \\
     0\,,            & x^\pm \geq 0.
  \end{array}\right.
\end{eqnarray}
Integrating twice, we obtain
\begin{eqnarray}\label{suddenr}
 r_\pm=
  \left\{
\begin{array}{ll}
2 \lambda^2 (x^\pm )^2 + \frac{a}{2}\,, &x^\pm < 0\\
   \frac{a}{2}\,, &x^\pm \geq 0
  \end{array}\right.
\end{eqnarray}
where the constants of integration are determined by continuity at $x^\pm = 0$.
The resulting solution from (\ref{rsolution}) is
\begin{eqnarray}\label{suddensol}
  r =
  \left\{
\begin{array}{ll}
    a + 2 \lambda^2 (x^+ - x^- )^2\,,    &x^+ < 0  ~{\rm and}~ x^- < 0\\
     a + 2 \lambda^2 x^+ (x^+ - 2x^- )\,, &x^+ < 0 ~{\rm and}~ x^- \geq 0 \\
     a - 2 \lambda^2 x^- (2x^+ - x^- )\,, & x^+ \geq 0 ~{\rm and}~ x^- < 0 \\
     a - 4 \lambda^2 x^+ x^-\,,       &x^+ \geq 0  ~{\rm and}~ x^- \geq 0.
  \end{array}\right.
\end{eqnarray}
One may recognize the solution in the final region as a static black hole
(\ref{vacuumsol}) with mass $m=a/2$. The double trapping horizon of the static
wormhole bifurcates into the event horizons of the black hole, all with
$r=a=2m$ (Fig.~\ref{sud}(b)). Thus the wormhole immediately collapses into a
black hole.

The Schwarzschild-like relationship between mass and throat radius is no
accident; there is a definition of active gravitational mass-energy \cite{cc}
\begin{equation}\label{energy}
  E=\frac{r}2\left(1-\frac{\nabla r\cdot\nabla r}{4\lambda^2r^2}\right)
\end{equation}
which evaluates as $r/2$ on any trapping horizon. Thus a wormhole with a throat
$r=a$ has an effective mass $a/2$. Here and elsewhere, it is useful to recall
the analogy with spherically symmetric Einstein gravity, where $r$ corresponds
to areal radius and there is a similar definition of active gravitational
mass-energy \cite{1st}.

\subsection{Gradual collapse}

\begin{figure}
\includegraphics[width=8cm,height=13cm,angle=0]{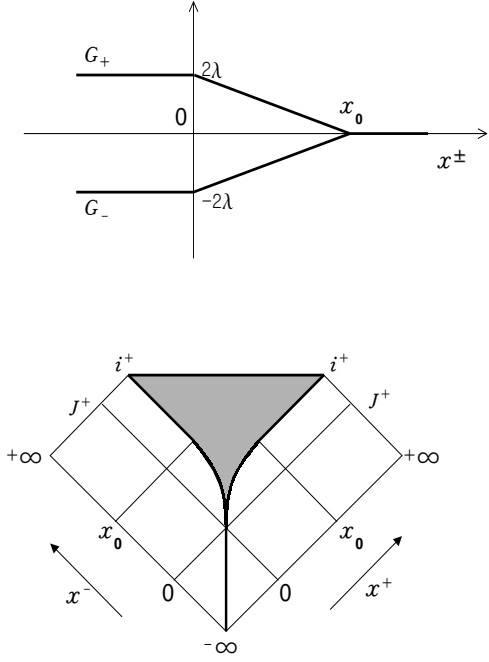}
 \vspace{-32mm}
 \caption{(a) The ghost field $g$ is gradually reduced to zero.
 (b) Conformal diagram: the wormhole throat bifurcates and the resulting
 non-static wormhole again eventually becomes a black hole.
 Shading in these diagrams indicates trapped regions,
 where $\partial_+r\partial_-r>0$.}
 \label{grad}
\end{figure}

Again starting with a static wormhole, we now reduce the ghost field gradually
to zero in the simplest way, linearly (Fig.~\ref{grad}(a)):
\begin{eqnarray}\label{gradualGeq}
  G_\pm=
   \left\{
 \begin{array}{ll}
   \pm 2 \lambda\,,   &x^\pm < 0\\
     \mp \alpha x^\pm \pm 2 \lambda\,, &0 \leq x^\pm < x_0 \\
     0\,, &x_0 \leq x^\pm
  \end{array} \right.
\end{eqnarray}
where $\alpha$ is a constant and $x_0 = 2\lambda/\alpha$. Here $\alpha\to0$
recovers the static wormhole and $\alpha\to\infty$ recovers sudden collapse. By
similar calculations, again taking $f=0$,
\begin{eqnarray}\label{gradualreq}
  \partial_\pm   \partial_\pm  r_\pm =
\left\{
  \begin{array}{ll}
     4\lambda^2\,, &x^\pm < 0\\
     \alpha^2 {x^\pm}^2 - 4 \lambda \alpha x^\pm + 4\lambda^2\,,
    &0 \leq x^\pm < x_0 \\
    0\,,  &x_0 \leq x^\pm
  \end{array}\right.
\end{eqnarray}
integrate to
\begin{eqnarray}\label{gradualr}
 r_\pm=
 \left\{
 \begin{array}{ll}
    2\lambda^2 {x^\pm}^2  + \frac{a}{2}\,, &x^\pm < 0\\
     \frac{\alpha^2}{12} {x^\pm}^4 -  \frac{2\alpha\lambda}{3}  {x^\pm}^3 + 2\lambda^2 {x^\pm}^2
     +  \frac{a}{2}\,,
       &0 \leq x^\pm < x_0 \\
      \frac{8 \lambda^3}{3\alpha} x^\pm -  \frac{4\lambda^4}{3\alpha^2}  +
      \frac{a}{2}\,,
     &x_0 \leq x^\pm.
  \end{array}\right.
\end{eqnarray}
and the solution
\begin{widetext}
\begin{eqnarray}\label{gradualsol}
  r=
  \left\{
 \begin{array}{ll}
   a +  2 \lambda^2 (x^+ - x^- )^2\,,
       & x^+ < 0 ~{\rm and}~ x^- < 0\\
      a +  2 \lambda^2 (x^+ - x^- )^2  -  \frac{2\alpha\lambda}{3} {x^- }^3
          +  \frac{\alpha^2}{12} {x^- }^4\,,
        &x^+ < 0 ~{\rm and}~ 0 \leq x^- < x_0 \\
    a +  2 \lambda^2(x^+ - x^- )^2  -  \frac{2\alpha\lambda}{3} {x^+ }^3
         +  \frac{\alpha^2}{12} {x^+ }^4\,,
      &0 \leq  x^+ < x_0 ~{\rm and}~  x^- < 0 \\
      a -  \frac{4\lambda^4}{3\alpha^2} - 4 \lambda^2 x^+ x^-
         +  \frac{8\lambda^3}{3\alpha} x^- + 2\lambda^2 {x^+} ^2\,,
           &x^+ < 0  ~{\rm and}~ x_0 \leq x^- \\
     a +  2 \lambda^2 (x^+ - x^- )^2  -  \frac{2\alpha\lambda}{3} ( {x^+}^3 + {x^- }^3)
       +  \frac{\alpha^2}{12}  ( {x^+}^4 + {x^- }^4)\,,
           &0 \leq x^+ < x_0 ~{\rm and}~ 0 \leq x^- <  x_0  \\
      a  -  \frac{4\lambda^4}{3\alpha^2} - 4 \lambda^2 x^+ x^-
         +  \frac{8\lambda^3}{3\alpha} {x^+ } + 2\lambda^2 {x^-
         }^2\,,
       & x_0 \leq x^+ ~{\rm and}~ x^- < 0\\
      a  - \frac{4\lambda^4}{3\alpha^2} - 4 \lambda^2 x^+ x^-  + \frac{8\lambda^3}{3\alpha} x^-
        + 2\lambda^2 {x^+ }^2
     -  \frac{2\alpha\lambda}{3} {x^+ }^3 +  \frac{\alpha^2}{12} {x^+
     }^4\,,
        &0 \leq x^+ <  x_0 ~{\rm and}~ x_0 \leq x^-\\
     a  -  \frac{4\lambda^4}{3\alpha^2} - 4 \lambda^2 x^+ x^-  +  \frac{8\lambda^3}{3\alpha} x^+
         + 2\lambda^2 {x^- }^2
       -  \frac{2\alpha\lambda}{3} {x^- }^3 +  \frac{\alpha^2}{12} {x^-
       }^4\,,
        & x_0 \leq x^+  ~{\rm and}~ 0 \leq x^- <  x_0 \\
      a  -  \frac{8\lambda^4}{9\alpha^2} - 4 \lambda^2 \left(x^+ -  \frac{2\lambda}{3\alpha} \right)
                                \left(x^-  -   \frac{2\lambda}{3\alpha}
                                \right)\,,
           & x_0 \leq x^+  ~{\rm and}~  x_0 \leq x^-.
  \end{array}\right.
 \end{eqnarray}
\end{widetext}

The solution in the final region is again recognizable as a black hole
(\ref{vacuumsol}), but with reduced mass and shifted by
$\frac{2\lambda}{3\alpha}=\frac{x_0}3$ in $x^\pm$ compared with the sudden
case. The trapping horizons $\partial_\pm r=0$ are located at
\begin{eqnarray}\label{gr-hor}
  x^\mp=
   \left\{
 \begin{array}{ll}
   x^\pm\,,   &x^\pm < 0\\
     x^\pm - \frac{1}{x_0}{x^\pm}^2
     +\frac{1}{3x_0^2}{x^\pm}^3\,, &0 \leq x^\pm < x_0 \\
     \frac{x_0}{3}\,, &x_0 \leq x^\pm.
  \end{array} \right.
\end{eqnarray}
The final expressions coincide with the locations of the event horizons of the
black hole. The middle expressions are cubic curves which smoothly join the
initial wormhole throat to the final event horizons (Fig.~\ref{grad}(b)). Thus
the double trapping horizon of the wormhole has again bifurcated,  eventually
forming a black hole. In between, the geometry may be characterized as a
non-static wormhole, as observers may still traverse it for a certain time,
after which the attempt leads only into the black hole.

\section{Wormhole construction}

\begin{figure}
\includegraphics[width=7cm,height=10cm,angle=0]{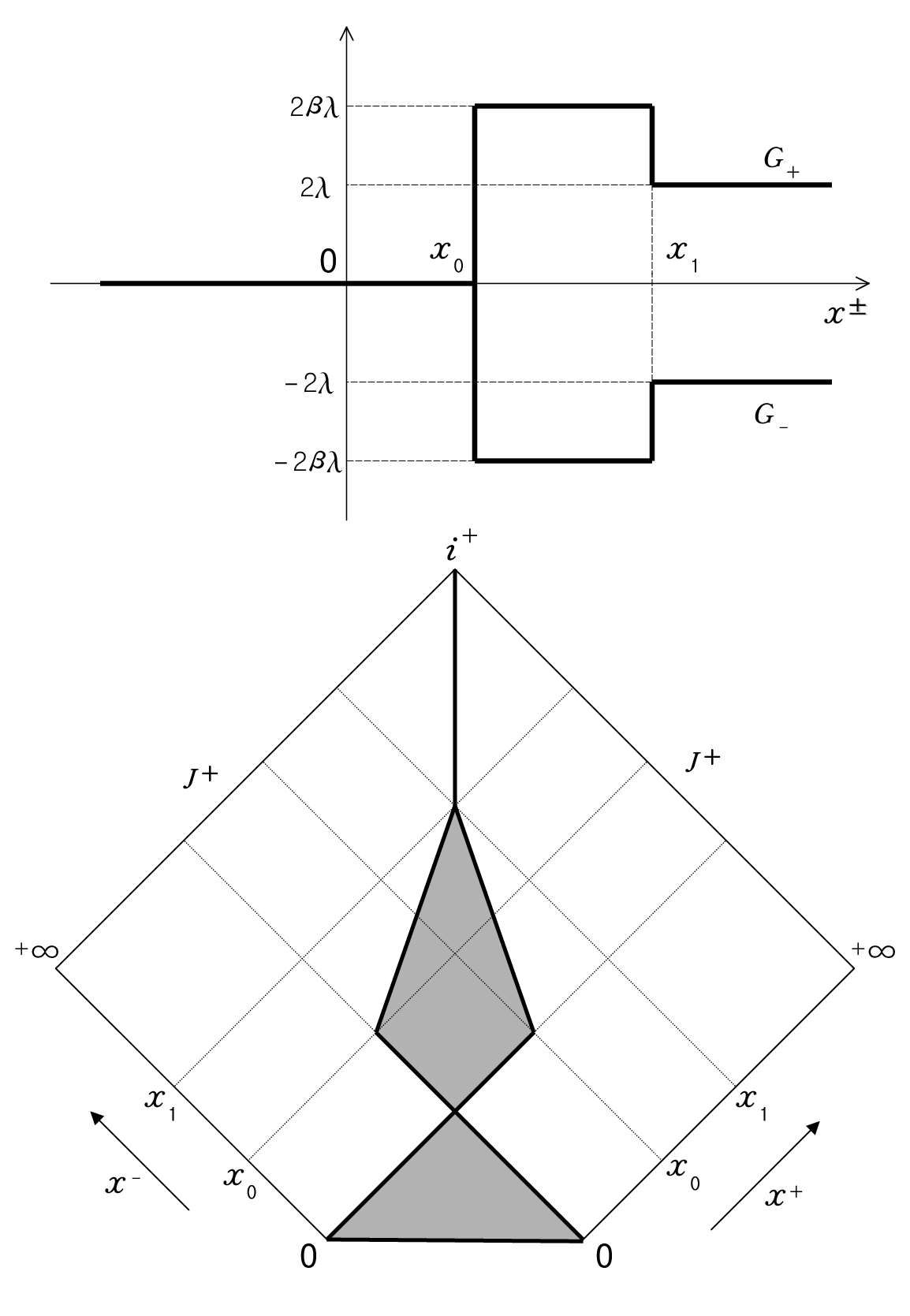}
 \vspace{-2mm}
 \caption{(a) Irradiating a vacuum black hole with the ghost field $g$.
 (b) Conformal diagram: the initially static black hole, as in
(Fig.~\ref{staticbhwh}(a)), becomes a dynamic wormhole, eventually reaching a
static state as in (Fig.~\ref{staticbhwh}(b)). The black hole has been
converted into a traversible wormhole.}
 \label{cons}
\end{figure}

We next study how to convert a black hole into a traversible wormhole by
irradiating it with the ghost field. This is not simply the time reverse of the
wormhole collapses described above, which would represent the creation of a
wormhole from a white hole, as one can see by inverting the conformal diagrams
(Fig.~\ref{sud}(b),\ref{grad}(b)). Instead, one needs to begin the irradiation
at some positive value $x_0$ of the Kruskal-like coordinates $x^\pm$
(Fig.~\ref{staticbhwh}(a)). Moreover, in order to close up a future trapped
region by merging its trapping horizons, the negative energy densities required
must be larger than those required to maintain the static wormhole which
finally forms \cite{wh}. In other words, $|G_\pm|$ must first reach a maximum
greater than $2\lambda$. The simplest way to achieve this is to set initial
data as double step functions (Fig.~\ref{cons}(a))
\begin{eqnarray}\label{constructGeq}
  G_\pm =
  \left\{
 \begin{array}{ll}
  0\,,  &x^\pm < x_0 \\
  \pm 2\beta\lambda\,, &x_0 \leq x^\pm < x_1 \\
  \pm 2 \lambda\,,  &x_1 \leq x^\pm
  \end{array}\right.
\end{eqnarray}
where $\beta>1$ is a constant, to be determined by the requirement that the
trapping horizons merge at $x^\pm=x_1$, for which we find
$x_0=x_1(1-\beta^{-2})$. Then taking $f=0$, the constraints (\ref{constraints})
are
\begin{eqnarray}\label{constructreq}
  \partial_\pm   \partial_\pm  r_\pm =
  \left\{
 \begin{array}{ll}
    0\,, &x^\pm <  x_0 \\
    4 \lambda^2\beta^2\,,
    & x_0 \leq x^\pm <  x_1 \\
    4 \lambda^2\,,  & x_1 \leq x^\pm.
  \end{array}\right.
\end{eqnarray}
Integrating twice, assuming a black hole of mass $m$ in the initial region,
leads to
\begin{eqnarray}\label{constructr}
 r_\pm =
  \left\{
 \begin{array}{ll}
    m \,,   &x^\pm <  x_0 \\
    2\lambda^2\beta^2(x^\pm-x_0)^2 + m\,,  &x_0 \leq x^\pm < x_1\\
    2 \lambda^2 ({x^\pm}^2  - x_0 x_1) + m\,,     &x_1 \leq x^\pm
  \end{array}\right.
\end{eqnarray}
and
\begin{widetext}
\begin{eqnarray}\label{constructsol}
  r=
 \left\{
 \begin{array}{ll}
  2m - 4\lambda^2  x^+ x^- \,,
      &x^+ <  x_0 ~{\rm and}~  x^- <  x_0 \\
  2m - 4\lambda^2 x^+ x^- + 2\lambda^2\beta^2\left((x^+-x_0)^2+(x^--x_0)^2\right)\,,
         & x_0 \leq x^+ <  x_1 ~{\rm and}~  x_0 \leq x^- <  x_1 \\
  2\lambda^2 (x^+ - x^-)^2 + 2m - 4\lambda^2 x_0 x_1 \,,
        & x_1 \leq x^+  ~{\rm and}~  x_1 \leq x^-
  \end{array}\right.
\end{eqnarray}
\end{widetext}
where we have omitted the less relevant regions. One may recognize the solution
in the final region as a static wormhole (\ref{staticwhsol}) with throat radius
$2m-4\lambda^2x_0x_1$. Thus we require $2\lambda^2x_0x_1<m$. By choice of the
parameters $(x_0,x_1)$, we have constructed a static wormhole with any throat
radius less than $2m$, the radius of the original black hole. The trapping
horizons $\partial_\pm r=0$ are located at
\begin{eqnarray}\label{construct-hor}
  x^\mp=
   \left\{
 \begin{array}{ll}
   0\,,   &x^\pm < x_0\\
   \beta^2(x^\pm-x_0) \,, &x_0 \leq x^\pm < x_1 \\
   x^\pm\,, &x_1 \leq x^\pm
  \end{array} \right.
\end{eqnarray}
which are straight line segments. Thus the ghost radiation causes the trapping
horizons of the initial black hole to shrink towards each other, eventually
merging to form the throat of the final static wormhole (Fig.~\ref{cons}(b)).
The trapped region composing the black hole simply evaporates. This classically
unexpected behaviour is, of course, due to the negative energy densities. As in
the wormhole collapse case, the trapping horizons can be smoothed off by taking
smoother profiles for $G_\pm$, but the details are not particularly
illuminating. In summary, static wormholes have been constructed by irradiating
a black hole with ghost radiation.

One can also regard this as analogous to black-hole evaporation, with the ghost
radiation modelling the ingoing negative-energy Hawking radiation, suggesting
that the final state of black hole evaporation might be a stationary wormhole
\cite{wh,HPS,KL}. This would naturally resolve the information-loss puzzle, as
there is no singularity in which information is lost; everything that fell into
the black hole eventually re-emerges.

\section{Wormhole operation}

If a wormhole is actually used to transport a parcel or person between the two
universes, the transported matter will affect the wormhole by changing the
gravitational field. In principle this occurs even if the wormhole is merely
used for signalling. We will study the dynamical effects of such back-reaction
by using the field $f$ to model the matter or radiation. The use of
Klein-Gordon radiation rather than more realistic matter is for simplicity
only; any source of mass-energy would have more or less similar gravitational
effects. In the current model, the constraints (\ref{constraints}) show
explicitly that increasing ${F_\pm}^2$ has an equivalent gravitational effect
to reducing ${G_\pm}^2$, which we have already shown can cause collapse to a
black hole. Thus it is to be expected that too much transport would destroy the
wormhole. A worse possibility is that the wormhole might be unstable to the
slightest perturbation and start to collapse immediately. We investigate this
below, giving the first concrete examples of wormhole operation including
back-reaction.

\begin{figure}
\includegraphics[width=8cm,height=11cm,angle=0]{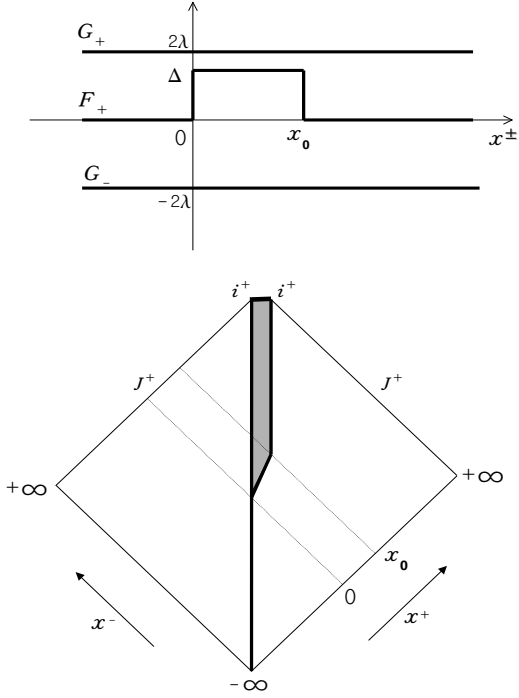}
 \vspace{-7mm}
 \caption{(a) A step pulse of positive-energy radiation is beamed
through the wormhole.
 (b) Conformal diagram: the wormhole becomes non-static but, for a
small-energy pulse, remains traversible for a long time.}
 \label{opr1}
\end{figure}

Again taking the simplest case, we consider a step pulse of positive-energy
radiation (Fig.~\ref{opr1}(a)):
\begin{eqnarray}\label{step}
  F_+ &=&
 \left\{
 \begin{array}{ll}
  0\,,~ &x^+ <  0 \\
    \Delta\,,& 0 \leq x^+ <  x_0 \\
    0\,,   & x_0 \leq x^+
  \end{array}\right. \\
F_- &=& 0 \nonumber
\end{eqnarray}
with $G_\pm = \pm 2\lambda$. The energy of the pulse may be defined as the
change in the gravitational energy (\ref{energy}) due to the pulse, which
evaluates at $x^-=0$ as
\begin{equation}
\epsilon=\frac14(\Delta x_0 )^2.
\end{equation}
In the following, we assume that the energy of the pulse should not be too
large, corresponding to the anticipated limit on how much mass-energy can be
sent through the wormhole without causing it to collapse into a black hole.
Specifically we find $\epsilon<a/2$ to avoid an $r=0$ singularity in the middle
region. Thus the pulse energy should be less than the effective mass of the
wormhole. Inserting the conversion factor $c^2/G$ from length to mass, a
one-metre wormhole could transport over a hundred Earth masses. This is hardly
a practical limitation, once we establish stability.

The constraints (\ref{constraints})
\begin{eqnarray}
\partial_+  \partial_+ r_+ &=&
  \left\{
 \begin{array}{ll}
    4\lambda^2\,, &x^+ <  0 \\
    4\lambda^2 - \Delta^2\,,& 0 \leq x^+ <  x_0 \\
    4 \lambda^2\,,   & x_0 \leq x^+
  \end{array}\right. \\
\partial_-  \partial_- r_- &=&4 \lambda^2\nonumber
\end{eqnarray}
integrate to
\begin{widetext}
\begin{eqnarray}
 r_+ &=&
  \left\{
 \begin{array}{ll}
    2\lambda^2 {x^+}^2 + \frac{a}{2}\,, &x^+ < 0 \\
   2\lambda^2 {x^+}^2 - \frac{\Delta^2}{2}{x^+}^2
   + \frac{a}{2}\,,~ & 0 \leq x^+ <  x_0 \\
   2\lambda^2 {x^+}^2 + \Delta^2 x_0 \left( \frac{1}{2}x_0 - x^+ \right)
   + \frac{a}{2}\,,
   ~& x_0 \leq x^+
  \end{array}\right.\\
 r_- &=&2\lambda^2 {x^-}^2 + \frac{a}{2}\nonumber
\end{eqnarray}
and the solution is
\begin{eqnarray}
 r =
  \left\{
 \begin{array}{ll}
    a + 2\lambda^2 \left({x^+} - {x^-} \right)^2\,, &x^+ <  0 \\
   a + 2\lambda^2 \left({x^+} - {x^-} \right)^2- \frac{\Delta^2}{2}{x^+}^2
   \,,~ & 0 \leq x^+ <  x_0 \\
   a + 2\lambda^2 \left({x^+} - {x^-} \right)^2
   + \Delta^2 x_0 \left( \frac{1}{2}x_0 - x^+ \right) \,,   ~& x_0 \leq x^+.
  \end{array}\right.\,
\end{eqnarray}
\end{widetext}
The locations of the trapping horizons $\partial_+r=0$ and $\partial_-r=0$ are
given respectively by
\begin{eqnarray}
x^- &=&
  \left\{
 \begin{array}{ll}
   x^+ \,, &x^+ < 0 \\
   x^+ - \frac{\Delta^2}{4\lambda^2} x^+
   \,,~ & 0 \leq x^+ <  x_0 \\
   x^+ - \frac{\Delta^2}{4\lambda^2} x_0
    \,,   ~& x_0 \leq x^+
  \end{array}\right.\\
x^+ &=& x^- \,.\nonumber
\end{eqnarray}
Thus the double trapping horizon of the initially static wormhole bifurcates
when the radiation arrives (Fig.~\ref{opr1}(b)). After the pulse has passed,
the two trapping horizons run parallel in the $x^\pm$ coordinates, forming a
non-static traversible wormhole. If the pulse energy is small, $\epsilon\ll a$,
the wormhole persists in an almost static state for a long time, $t\sim
x_0a/\epsilon$. Nevertheless, even for an arbitrarily weak pulse, eventually a
spatial $r=0$ singularity develops, similar to that of the static black hole.
Observers close enough to the singularity can no longer traverse the wormhole,
so it constitutes a black hole with two event horizons. Thus the static
wormhole exhibits a type of neutral stability, neither strictly stable in that
it does not return to its initial state, nor strictly unstable in that there is
no sudden runaway. Note that black holes are also neutrally stable in this
sense; perturbing a black hole by dropping positive-energy matter into it
increases the area of the trapping horizon finitely, by the first and second
laws of black-hole dynamics \cite{cc,1st,bhd}.

\section{Wormhole maintenance}

Keeping an operating wormhole viable indefinitely, defying its natural fate as
a black hole, requires additional negative energy to balance the transported
matter. The simplest way to maintain the wormhole is just to set initial data
\begin{equation}
 G_\pm = \pm\sqrt{4\lambda^2 + {F_\pm}^2}
\end{equation}
so that the source terms in the constraints (\ref{constraints}) cancel to the
static wormhole values. Thus the wormhole remains static. The additional
negative-energy radiation has balanced the positive-energy radiation, leaving
the gravitational field unchanged.

\begin{figure}
\includegraphics[width=7cm,height=10cm,angle=0]{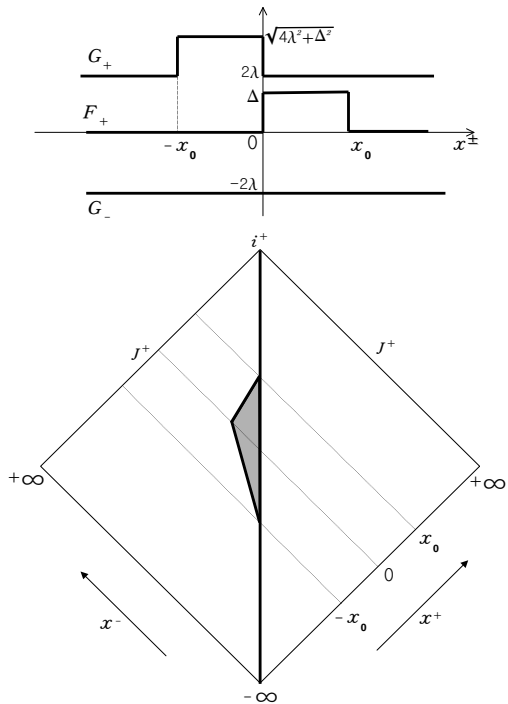}
 \caption{(a) The step pulse of positive-energy radiation is balanced by a
preceding pulse of negative-energy radiation.
 (b) Conformal diagram: the wormhole returns to its original static state.}
 \label{opr2}
\end{figure}

Alternatively, the wormhole may be maintained by beaming in additional
negative-energy radiation before or after the positive-energy pulse. We take
the same step pulse in $F_+$ (\ref{step}) and precede it with a compensating
pulse in the ghost field (Fig.~\ref{opr2}(a)):
\begin{eqnarray}\label{addG}
  G_+ &=&
 \left\{
 \begin{array}{ll}
  2\lambda\,,~ &x^+ < -x_0 \\
  \sqrt{4\lambda^2 + \Delta^2}\,,~ & -x_0 \leq x^+ < 0  \\
  2\lambda\,,   & 0 \leq x^+
  \end{array}\right.\,\\
 G_- &=& -2\lambda\, .\nonumber
\end{eqnarray}
Again we require small pulse energy, $\epsilon<a/2$, to avoid a singularity.
The constraints (\ref{constraints}) become
\begin{eqnarray}
\partial_+ \partial_+ r_+ &=&
  \left\{
 \begin{array}{ll}
    4\lambda^2\,, &x^+ < -x_0 \\
    4\lambda^2 + \Delta^2\,,& -x_0 \leq x^+ < 0 \\
    4\lambda^2 - \Delta^2\,,& 0 \leq x^+ < x_0 \\
    4 \lambda^2\,,   & x_0 \leq x^+
  \end{array}\right. \\
\partial_-  \partial_- r_- &=&4 \lambda^2\nonumber
\end{eqnarray}
which integrate to
\begin{widetext}
\begin{eqnarray}
  r_+ &=&
 \left\{
 \begin{array}{ll}
  2\lambda^2 {x^+}^2 + \frac{a}{2}\,, &x^+ < -x_0 \\
   2\lambda^2 {x^+}^2 + \Delta^2 x^+ \left(\frac{1}{2}x^+ + x_0 \right)
   -\frac{1}{2}\Delta^2 {x_0}^2 + \frac{a}{2} \, ,
        ~ & -x_0 \leq x^+ < 0 \\
   2\lambda^2 {x^+}^2 - \Delta^2 x^+ \left(\frac{1}{2}x^+ - x_0 \right)
   -\frac{1}{2}\Delta^2 {x_0}^2 + \frac{a}{2} \, ,
        ~ & 0 \leq x^+ <  x_0 \\
  2\lambda^2 {x^+}^2 + \frac{a}{2}\,, &x_0 \leq x^+
  \end{array}\right.\\
r_- &=&2\lambda^2 {x^-}^2 + \frac{a}{2}\nonumber
\end{eqnarray}
and the solution follows as
\begin{eqnarray}
r=
 \left\{
 \begin{array}{ll}
  a + 2\lambda^2 \left({x^+} - {x^-} \right)^2\,, & x^+ < -x_0 \\
   a + 2\lambda^2 \left({x^+} - {x^-} \right)^2
   + \frac{1}{2}\Delta^2 \left({x^+}^2 + 2x_0 x^+ - {x_0}^2 \right) \,,
        ~ & -x_0 \leq x^+ < 0 \\
   a + 2\lambda^2 \left({x^+} - {x^-} \right)^2
   - \frac{1}{2}\Delta^2 \left({x^+}^2 - 2x_0 x^+ + {x_0}^2 \right) \,,
          ~& 0 \leq x^+ < x_0 \\
  a + 2\lambda^2 \left({x^+} - {x^-} \right)^2 \,,   ~& x_0 \leq x^+.
  \end{array}\right.
\end{eqnarray}
\end{widetext}
The locations of the trapping horizons are
\begin{eqnarray}
x^- &=&
 \left\{
 \begin{array}{ll}
  x^+ \,, &x^+ < -x_0 \\
  x^+ + \frac{\Delta^2}{4\lambda^2}\left( x^+ + x_0 \right) \,,
        ~ & -x_0 \leq x^+ < 0 \\
  x^+ - \frac{\Delta^2}{4\lambda^2}\left( x^+ - x_0 \right) \,,
          ~& 0 \leq x^+ < x_0 \\
  x^+ \,, ~& x_0 \leq x^+
  \end{array}\right.\\
x^+ &=& x^- \nonumber
\end{eqnarray}
which again are straight line segments. The double horizon bifurcates when the
ghost pulse arrives, temporarily opening up a trapped region, but the two
horizons subsequently merge to form a static wormhole again
(Fig.~\ref{opr2}(b)). This is not unexpected, since the energy of the ghost
pulse, $\epsilon'=-\Delta^2{x_0}^2/4$, balances that of the other pulse:
\begin{equation}
\epsilon+\epsilon'=0.
\end{equation}
In fact, the final state is identical to the initial state in this symmetric
case. The wormhole can also be returned to a different static state, with
different throat radius, by less symmetric double pulses. The examples show
that there is no practical problem of fine-tuning the ghost field to keep the
wormhole static. An almost static wormhole is still traversible and can be
adjusted at any time to bring it closer to staticity, or to change its size.
This is essentially due to the neutral stability.

\section{Conclusion}

Space-time wormholes remain in the realms of science fiction and theoretical
physics. By the standards of either genre, they are not so far-fetched,
differing from experimentally established physics by only one step, the
existence of negative energy densities in sufficient concentrations. Here we
have not addressed this issue but, assuming a positive answer, have
investigated the behaviour of the resulting wormholes, evolving dynamically
according to field equations. In particular, we have found detailed answers to
the following practical questions. (i) How can one construct a traversible
wormhole? By bathing a black hole in exotic radiation. (ii) Is an operating
wormhole stable under the back-reaction of the transported matter? In this
case, neutrally stable. (iii) How can a wormhole be maintained indefinitely for
transport or signalling? By balance of positive and negative energy. (iv) What
happens if the negative-energy source fails? The wormhole collapses into a
black hole.

This was mostly predicted by a general theory of wormhole dynamics \cite{wh},
but here we have given concrete examples, by virtue of the exact solubility of
the field equations of two-dimensional dilaton gravity. Despite the supposedly
unphysical nature of a ghost Klein-Gordon field, fears about instability,
runaway processes and naked singularities proved to be unfounded. More
realistic situations may differ in some respects, such as wormhole stability,
which may be affected by backscattering and will presumably depend on the
exotic matter model. However, the same principles and methods should apply,
such as energy balance and tracking of trapping horizons. Indeed, apart from
the inclusion of exotic matter or radiation, the methods are the same as those
used to analyze black-hole dynamics \cite{1st,bhd}. In particular, the explicit
examples of dynamic interconversion of black holes and wormholes should assuage
objections that they are fundamentally different objects. Rather, wormholes and
black holes have similar physics and a unified theory.

\medskip\noindent
S.A.~Hayward was supported by Korea Research Foundation grant
KRF-2001-015-DP0095, S-W.~Kim by Korea Research Foundation grant
KRF-2000-041-D00128 and H.~Lee by KOSEF grant R01-2000-00015.

\end{document}